\documentclass[10pt,twoside,twocolumn,english,aps,prl,superscriptaddress,notitlepage]{revtex4-2}
\usepackage[utf8]{inputenc}
\usepackage[a4paper]{geometry}
\usepackage{color,xcolor,soul}
\usepackage{verbatim,textcomp}
\usepackage{amstext,amsfonts,amsmath,amssymb}
\usepackage{graphicx}
\usepackage[calcwidth,explicit]{titlesec}
\usepackage{chngcntr}
\usepackage[normalem]{ulem}
\usepackage[unicode=true,pdfusetitle, bookmarks=false, breaklinks=true,pdfborder={0 0 0},pdfborderstyle={},backref=false,colorlinks=true]{hyperref}
\hypersetup{colorlinks=true,citecolor=blue,linkcolor=blue,urlcolor=blue}

\makeatletter\renewcommand\frontmatter@abstractwidth{\dimexpr\textwidth-2cm\relax}\makeatother

\titleformat{\section}{}{}{0pt}{}
\titleformat{\section}{\bfseries\sffamily\large\filcenter}{\thesection.}{0.2em}{#1}
\titlespacing{\section}{0pt}{0.2ex}{0.2ex}
\titleformat{\paragraph}[runin]{\normalfont\normalsize\bfseries}{}{0pt}{}
\titlespacing*{\paragraph}{0em}{0ex}{0.5em}[]

{}

\setcitestyle{super}

\renewcommand\thesection{\Alph{section}}

\AtBeginDocument{
	\renewcommand{\ref}[1]{\autoref{#1}}

}

\begin{document}
\title{Tailoring Zero-Field Magnetic Skyrmions in Chiral Multilayers \\by a
Duet of Interlayer Exchange Couplings\smallskip{}
}
\author{Xiaoye Chen}\thanks{These authors contributed equally to this work}
\email{chen\_xiaoye@imre.a-star.edu.sg}
\affiliation{Institute of Materials Research \& Engineering, Agency for Science, Technology \& Research (A{*}STAR), 138634 Singapore}
\author{Tommy Tai}\thanks{These authors contributed equally to this work}
\affiliation{Institute of Materials Research \& Engineering, Agency for Science, Technology \& Research (A{*}STAR), 138634 Singapore}
\author{Hui Ru Tan}\thanks{These authors contributed equally to this work}
\affiliation{Institute of Materials Research \& Engineering, Agency for Science, Technology \& Research (A{*}STAR), 138634 Singapore}
\author{Hang Khume Tan}
\affiliation{Institute of Materials Research \& Engineering, Agency for Science, Technology \& Research (A{*}STAR), 138634 Singapore}
\author{Royston Lim}
\affiliation{Institute of Materials Research \& Engineering, Agency for Science, Technology \& Research (A{*}STAR), 138634 Singapore}
\author{Pin Ho}
\affiliation{Institute of Materials Research \& Engineering, Agency for Science, Technology \& Research (A{*}STAR), 138634 Singapore}
\author{Anjan Soumyanarayanan}
\email{anjan@imre.a-star.edu.sg}
\affiliation{Institute of Materials Research \& Engineering, Agency for Science, Technology \& Research (A{*}STAR), 138634 Singapore}\affiliation{Physics Department, National University of Singapore (NUS), 117551 Singapore}

\begin{abstract}
Magnetic skyrmions have emerged as promising elements for encoding information towards biomimetic computing applications due to their pseudoparticle nature and efficient coupling to spin currents.
A key hindrance for skyrmionic devices is their instability against elongation at zero magnetic field (ZF). 
Prevailing materials approaches focussed on tailoring skyrmion energetics have found ZF configurations to be highly sensitive, which imposes significant growth constraints and limits their device scalability. 
Here we propose that designer ZF skyrmion configurations can be robustly stabilized within chiral multilayer stacks by exploiting a duet of interlayer exchange couplings (IECs). 
Microscopic imaging experiments show that varying the two IECs enables the coarse and fine-tuning of ZF skyrmion stability and density.
Micromagnetic simulations reveal that the duo-IEC approach is distinguished by its influence on the kinetics of skyrmion nucleation, in addition to the ability to tailor energetics, resulting in a substantially expanded parameter space, and enhanced stability for individual ZF skyrmions.  
Our work underscores the importance of IEC as a means of stabilizing and controlling ZF skyrmions, paving the way to scalable skyrmion-based devices.
\end{abstract}
\maketitle

\section{Introduction\label{sec:Intro}}

\paragraph{Isolated Skyrmions for Racetracks}
The magnetic skyrmion is a nanoscale topological magnetic texture formed in chiral, multilayered films \citep{Woo.2016,MoreauLuchaire.2016,Boulle.2016,Soumyanarayanan.2017}. It is a promising candidate for low-power, biomimetic computing due to its ease of electrical manipulation \citep{Jiang.2015,Buttner.2017,Woo.2018,Je.2021} and detection \citep{Neubauer.2009,Nagaosa.2013}, especially within linear ``racetrack'' device architectures \citep{Bourianoff.2018,Grollier.2020,Song.2020}. In contrast to domain walls (DWs), which can only move along the track length \citep{Parkin.2008}, skyrmions have two spatial degrees of freedom, providing much greater flexibility for applications. However, similar to DWs, information is encoded in the presence or absence of skyrmions. Hence, spatially isolated skyrmions of tunable density are more relevant to such racetrack applications than denser ensembles of magnetic textures.

\paragraph{Motivation: ZF Skyrmions}
A crucial requirement for the use of skyrmions in functional devices is their stability in the absence of external magnetic fields. However, isolated skyrmions, typically formed at finite out-of-plane (OP) fields, are known to elongate at zero field (ZF) into magnetic stripes (i.e., DW pairs) \citep{MoreauLuchaire.2016,Woo.2016,Soumyanarayanan.2017}. Such ZF proliferation of DWs is a direct consequence of their low or negative energy relative to the uniform state \citep{Bogdanov.1994}. While ZF skyrmions can be generated by current pulses \citep{Lemesh.2018,Brock.2020,Ang.2020} or stabilized by field reversal protocols \citep{Tan.2020}, they are often interspersed among stripes, which limits their utility. Alternatively, one could employ geometric confinement to realize isolated ZF skyrmions \citep{Boulle.2016,Zeissler.2017,Ho.2019}, but this in turn limits their lateral transport.

\begin{figure}
\centering\includegraphics[width=0.85\linewidth]{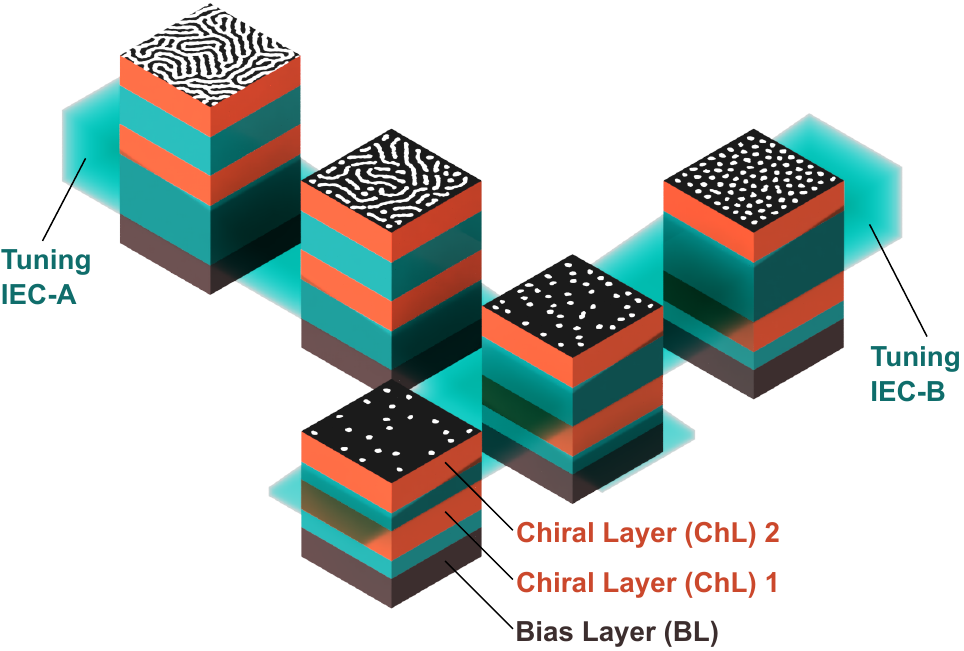}
\caption[Schematic Textural Tuning of ZF Skyrmions.]
{\textbf{Stability and density tuning of zero-field (ZF) skyrmions.} Schematic of the composite stack for tailoring ZF skyrmions, consisting of a bias layer (BL) and two chiral layers (ChL1,2). They are coupled by two interlayer exchange couplings (IEC-A,B), which act as coarse and fine `knobs' for ZF textural tuning.}
\label{fig:IEC-Schematic}
\end{figure}

\paragraph{Sk-IEC State-of-the-Art}
An emerging approach for materially stabilizing ZF skyrmions utilizes the interaction between a coercive magnetic layer (\textbf{\emph{bias layer, BL}}) and the skyrmion-hosting \textbf{\emph{chiral layer (ChL)}}\citep{Nandy.2016}, e.g., via interlayer exchange coupling (IEC), which involves the indirect Ruderman--Kittel--Kasuya--Yosida (RKKY) interaction between two magnetic layers mediated by a heavy-metal (HM) spacer \citep{Parkin.1991,Bruno.1995,Stiles.1999}. The IEC-mediated BL--ChL interaction can be simply conceptualized as an OP Zeeman field, $H_{\text{IEC}}$ acting on the ChL (\ref{fig:IEC-Schematic}: IEC-A)\textbf{ }\citep{Nandy.2016,Legrand.2019,LoConte.2020}, which can potentially realize skyrmions at ZF \citep{Chen.2015,Nandy.2016,LoConte.2020}. At present, such approaches are constrained by the high sensitivity of IEC to the spacer, which imposes epitaxial growth requirements, while limiting texture homogeneity and robustness. Alternative approaches utilizing exchange bias coupling of ChL to antiferromagnetic BLs \citep{Rana.2020,Guang.2020} require additional, intermediate magnetic layers, which may introduce material complexities. Notwithstanding these efforts, realizing isolated ZF skyrmions in device-compatible stacks may require mechanistic insights beyond thermodynamic stability.

\begin{figure*}[htb]
	\centering\includegraphics[width=0.9\linewidth]{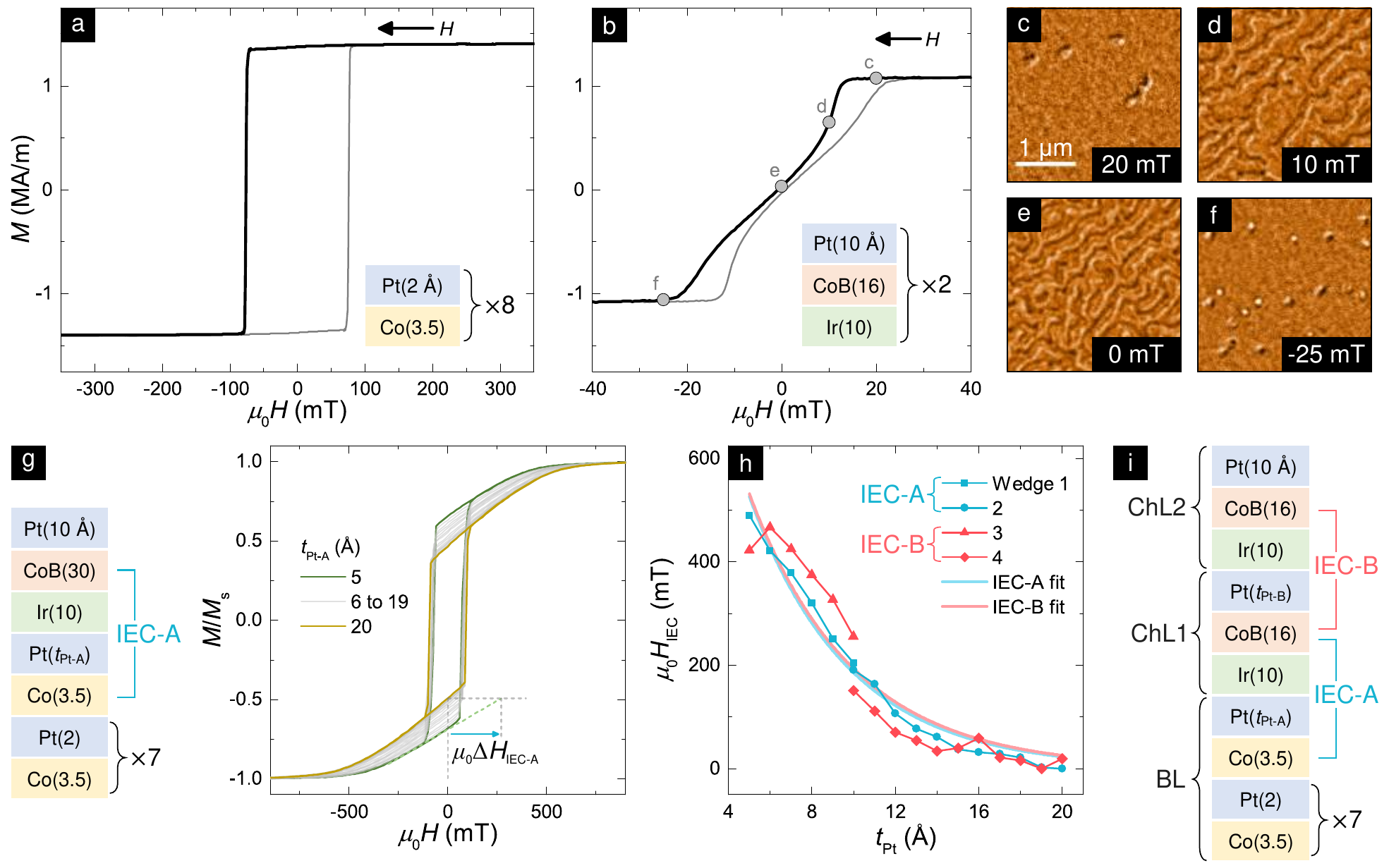}
	\caption[Characterization of Functional Layers.]{\textbf{Characterization of Functional Layers.} 
	Measured out-of-plane (OP) hysteresis loops of magnetization, $M(H)$, for \textbf{(a)} BL and \textbf{(b)} ChL bilayer. Insets depict individual stack structures.
	\textbf{(c-f)} LTEM-imaged textural evolution of ChL bilayer with applied OP field at selected fields (circles in (b)). 
	\textbf{(g)} Left: mixed-anisotropy stack structure used to determine IEC-A. Right:	Measured OP $M(H)$ loops for 16 studied samples with $t_{\text{Pt-A}}$ ranging from 5 to 20~Å. Extrapolated linear fits to these curves enables relative estimation of IEC-A, $\mu_{\text{0}}\Delta H_{\text{IEC-A}}$ (cyan arrow). 
	\textbf{(h)} Estimated actual IEC, $\mu_{\text{0}}H_{\text{IEC}}$, rescaled to CoB(16), extracted from two mixed-anisotropy stacks, representing IEC-A (red) and IEC-B (red) respectively (random error $\sim1\%$, not shown). Each stack is deposited via two wedged wafers (different symbols), with inter-wedge $t_{\text{Pt}}$ drift $\sim0.5$~Å. Solid lines are fits to an exponential decay form, $H_{\text{IEC}}\propto\exp(-t_{\text{Pt}}/t_{0})$. \textbf{(i)} The composite stack structure, comprising BL, Ch1, and ChL2, coupled pairwise by IEC-A and IEC-B respectively. }
	\label{fig:FuncLayers}
\end{figure*}

\paragraph{Results Summary}
In this work, we establish multiple causal links between IEC and ZF skyrmion stability. Our composite multilayer stack, comprising a BL and a ChL bilayer, exploits the interplay between two IEC knobs, tuned by the respective spacer thicknesses (\ref{fig:IEC-Schematic}). Lorentz transmission electron microscopy (LTEM) imaging show that while one knob controls the type of magnetic textures stabilized at ZF (\ref{fig:IEC-Schematic}: IEC-A) \citep{Chen.2015,Nandy.2016,LoConte.2020}, the other knob can independently fine-tune the \textbf{\emph{ZF skyrmion density, $n_{\text{s}}^{\text{Z}}$}} (\ref{fig:IEC-Schematic}: IEC-B). Micromagnetic simulations elucidate that this second IEC -- distinct from previous works -- influences the kinetics of ZF skyrmion nucleation. Importantly, the introduction of an extra IEC knob greatly enlarges the parameter space where individual ZF skyrmions are nucleated, and remain stable. Our findings advocate the use of IEC engineering to mold bespoke magnetic textures for device applications.

\section{Stack Design and Characterization\label{sec:StackDesign}}

\paragraph{Stack Design Considerations}
We begin by individually optimizing the two sputter-deposited functional layers -- BL and ChL -- before investigating their IEC-coupled behaviour. The BL is designed to be magnetically hard, similar to the pinned layer of a perpendicular magnetic tunnel junction \citep{Grimaldi.2020}. The optimized stack structure is {[}Co(3.5)/Pt(2){]}$_{8}$ (thickness in angstroms in parentheses), with coercive field, $\mu_{0}H_{{\rm c}}\sim75$~mT (\ref{fig:FuncLayers}a, details in SM §S1). Meanwhile, the ChL consists of an inversion-asymmetric trilayer, with a ferromagnet (FM) sandwiched by HMs to generate interfacial Dzyaloshinskii--Moriya interaction \citep{Bode.2007}. The ChL stack structure is Ir(10)/Co$_{80}$B$_{20}$(16)/Pt(10) (OP magnetization loop, $M(H)$ of ChL bilayer: \ref{fig:FuncLayers}b), wherein the CoB thickness is tuned to achieve a low effective anisotropy ($K_{\text{eff}}\gtrsim0$) for ease of hosting chiral magnetic textures (details in SM §S1)\citep{Soumyanarayanan.2017,Chen.2022}. To image magnetic textures in these ultrathin films, we use the well-established technique of full-field LTEM (see Methods, SM §S2) \citep{Phatak.2016,Fallon.2019,Garlow.2019}. \ref{fig:FuncLayers}c-f show the texture evolution along the upper branch (decreasing $H$) of the hysteresis loop. Isolated Néel skyrmions, approximately 150~nm in diameter, \citep{Chen.2022} are observed at around $\pm$20~mT (\ref{fig:FuncLayers}c,f) while the ZF configuration comprises labyrinthine stripes (\ref{fig:FuncLayers}e). Intuitively, skyrmions will be stabilized at ZF if the upper hysteretic branch (\ref{fig:FuncLayers}b: bold) is left-shifted by $\sim20$~mT, e.g. by introducing an IEC field.

\begin{figure*}
	\centering \includegraphics[width=1\linewidth]{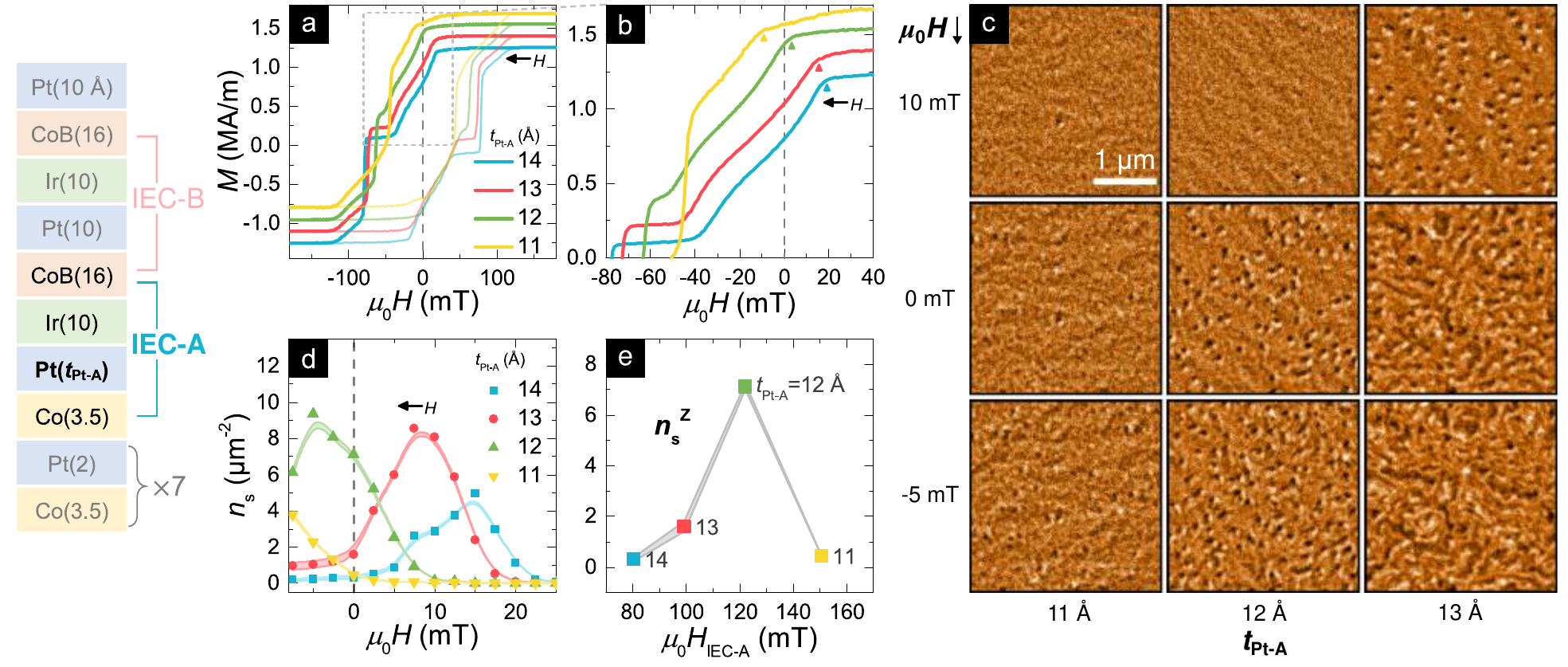}
	\caption[Varying $t_{\text{Pt-A}}$ (IEC-A).]{\textbf{Varying $t_{\text{Pt-A}}$ (IEC-A).} \textbf{(a)} OP $M(H)$ loops for the composite stack with $t_{\text{Pt-A}}$ varying over 11-14~~Å ($t_{\text{Pt-B}}=10$~Å). Each successive line plot is vertically offset by 0.15~MA/m for clarity. 
	\textbf{(b)} Zoom-in view of $M(H)$ loops (a, dotted box), showing the field evolution of the ChL bilayer. Arrows ($\blacktriangle$) mark the onset of the ChL bilayer evolution.
	\textbf{(c)} Representative LTEM images (scale bar: 1~$\mu$m, tilt angle: 15°, defocus $-2.4$~mm), illustrating the texture evolution with varying $t_{\text{Pt-A}}$ and decreasing external OP field $\mu_{0}H$. 
	\textbf{(d)} Field dependence of skyrmion density, $n_{\text{s}}(H)$, for $t_{\text{Pt-A}}$ varying from 11 - 14~Å. 
	\textbf{(e)} ZF skyrmion density, $n_{\text{s}}^{\text{Z}}$, plotted as a function of varying $H_{\text{IEC-A}}$ (obtained from $t_{\text{Pt-A}}$ via fit in \ref{fig:FuncLayers}h). Corresponding $t_{\text{Pt-A}}$ is indicated next to each data point.}
	\label{fig:IEC-A}
\end{figure*}

\paragraph{IEC Estimations}
We modulate the IEC between Co (or CoB) and CoB layers via the HM thicknesses (\ref{fig:FuncLayers}i). The thickness of Ir is fixed at 10~Å to maintain ferromagnetic IEC \citep{Parkin.1991}, while the thickness of Pt is varied ($t_{\text{Pt-A}}$ for IEC-A, $t_{\text{Pt-B}}$ for IEC-B) to tailor the IEC strength. Given the central role of ferromagnetic IEC variation to the results reported in this work, we have developed a \textbf{\emph{mixed-anisotropy magnetometry}} technique to estimate the two IEC magnitudes, validated by micromagnetic simulations (see SM §S3). In comparison to reported IEC estimation techniques \citep{Legrand.2019}, our approach does not require complex synthetic antiferromagnetic stacks, and also works for samples with high IEC. Instead, it utilises a mixed-anisotropy stack comprising an OP anisotropic BL and a thick in-plane (IP) anisotropic ``test'' CoB(30) layer, coupled across varying Pt thicknesses ($t_{\text{Pt}}$ = 5 - 20~Å; $t_{\text{Ir}}$ = 10~Å). Here, we outline the procedure for estimating IEC-A (\ref{fig:FuncLayers}g). As $t_{\text{Pt-A}}$ is reduced from 20~Å to 5~Å, the $H_{\text{IEC-A}}$ experienced by the IP layer increases. Acting similarly to an OP field, $H_{\text{IEC-A}}$ tilts IP magnetization (SM §S3), effectively increasing the measured OP magnetization (\ref{fig:FuncLayers}g). The relative change in $H_{\text{IEC-A}}$ between samples with $t_{\text{Pt-A}}$ of 5~Å and 20~Å, $\mu_{\text{0}}\Delta H_{\text{IEC-A}}$, is estimated by linearly fitting the hard axis component of the $M(H)$ loop for $t_{\text{Pt-A}}=5$~Å, and extrapolating it to the remanent magnetization of $t_{\text{Pt-A}}=20$~Å (\ref{fig:FuncLayers}g). This procedure is similarly repeated for all $t_{\text{Pt-A}}$ relative to $t_{\text{Pt-A}}=20$~Å. Next, we assume that for $t_{\text{Pt-A}}$=20~Å, IEC-A is negligible due to its rapid decay with thickness, allowing us to obtain the absolute value of IEC-A for each $t_{\text{Pt-A}}$. Finally, since $H_{\text{IEC}}$ is an effective field of interfacial origin, we rescale it by the inverse of the CoB layer thicknesses ratio (i.e., by 30/16) to obtain $\mu_{\text{0}}H_{\text{IEC-A}}$ as a function of $t_{\text{Pt-A}}$ for the ChL in \ref{fig:FuncLayers}h. A similar stack and methodology is used to obtain $\mu_{\text{0}}H_{\text{IEC-B}}$ (see SM §S3). Both IEC fields fit well to an exponential decay form, $H_{\text{IEC}}\propto\exp(-t_{\text{Pt}}/t_{0})$, where $t_{0}$ is 4.8$\pm0.3$~Å for IEC-A and 4.9$\pm0.6$~Å for IEC-B respectively. The decay length scale, $t_{0}$, is in line with previous reports \citep{Legrand.2019} notwithstanding differences in FM and HM layers. These fits are used subsequently to convert $t_{\text{Pt}}$ to $H_{\text{IEC}}$.

\paragraph{Composite Stack Design}
The composite stack thus established (\ref{fig:FuncLayers}i) comprises a BL coupled to a ChL (ChL1) via IEC-A ($t_{\text{Pt-A}}$), which is in turn coupled to another ChL (ChL2) via IEC-B ($t_{\text{Pt-B}}$). Notably, ChL2 is not directly coupled to the BL, but indirectly via ChL1. Hence, the effective field experienced by ChL2 will be weaker than that by ChL1. Further, we also expect some competition between IEC-A and IEC-B as both are coupled to ChL1, and the ensuing magnetic textures will likely represent the interplay between these two interactions. In the following section, we explore how varying IEC-A and IEC-B modifies the $M(H)$ loops and the ZF skyrmion density, $n_{\text{s}}^{\text{Z}}$.

\section{IEC Tuning\label{sec:IECTuning-Expts}}

\begin{figure*}
	\centering \includegraphics[width=1\linewidth]{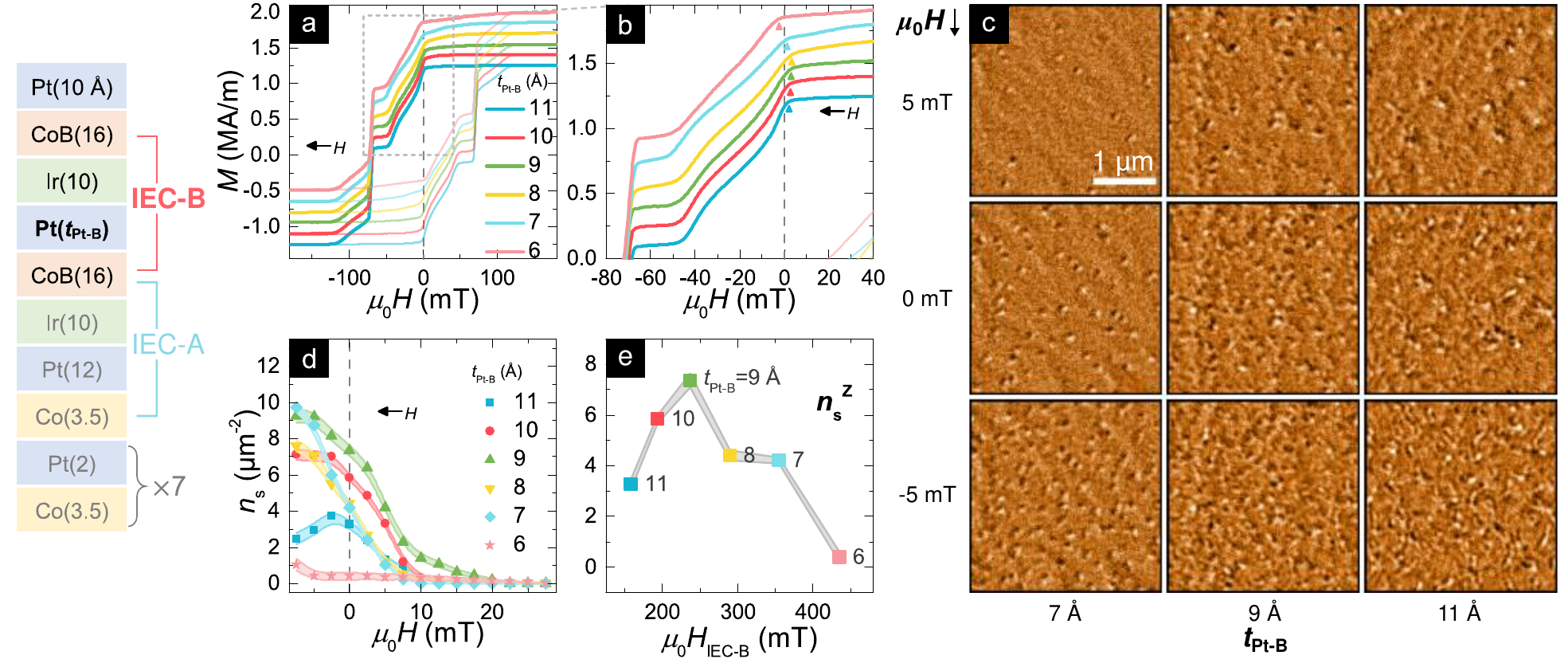} 
	\caption[Varying $t_{\text{Pt-B}}$ (IEC-B)]{\textbf{Varying $t_{\text{Pt-B}}$ (IEC-B).} \textbf{(a)} OP $M(H)$ loops for the composite stack with $t_{\text{Pt-B}}$ varying over 6-11~Å ($t_{\text{Pt-A}}=12$~Å). Each successive line plot is vertically offset by 0.15~MA/m for clarity. 
	\textbf{(b)} Zoom-in view of $M(H)$ loops (a: dotted box), showing the field evolution of the ChL bilayer (onset marked by arrows, $\blacktriangle$). 
	\textbf{(c)} Representative LTEM images (scale bar: 1~$\mu$m) illustrating the texture evolution with varying $t_{\text{Pt-B}}$ and decreasing $\mu_{0}H$.
	\textbf{(d)} Plots of skyrmion density, $n_{\text{s}}(H)$, for $t_{\text{Pt-B}}$ varying over 6 -- 11~Å. 
	\textbf{(e)} ZF skyrmion density, $n_{\text{s}}^{\text{Z}}$, plotted as a function of $H_{\text{IEC-B}}$ (obtained from $t_{\text{Pt-B}}$ via fit in \ref{fig:FuncLayers}h). Corresponding $t_{\text{Pt-B}}$ is indicated next to each data point.}
	\label{fig:IEC-B}
\end{figure*}

\paragraph{IEC-A: MH loop}
\ref{fig:IEC-A}a,b show the $M(H)$ loops of the composite stack with varying $t_{\text{Pt-A}}$ and fixed $t_{\text{Pt-B}}=10$~Å. We begin by consider the upper branch (decreasing $H$, in bold) for $t_{\text{Pt-A}}=14$~Å sample (blue, weakest IEC-A). The relatively weak interaction between the BL and the ChL bilayer results in two distinct segments in the $M(H)$ curve, with each resembling a constituent stack (\ref{fig:IEC-A}b: blue c.f. \ref{fig:FuncLayers}a-b). The field region from $+20$ to $-40$~mT is similar to the ChL bilayer field evolution (\ref{fig:FuncLayers}b), from uniform magnetisation (UM) to isolated skyrmions, to stripes, and then finally to skyrmions and UM of the opposite polarity (\ref{fig:FuncLayers}c-f). The sharp drop at $-75$~mT mirrors the BL switching (\ref{fig:FuncLayers}a). IEC-A acts as an effective field on the ChL bilayer ($H_{\text{IEC-A}}$), and decreasing $t_{\text{Pt-A}}$ increases $H_{\text{IEC-A}}$, which in turn left-shifts the ChL bilayer field evolution. \ref{fig:IEC-A}b shows the onset of the ChL bilayer field evolution (\ref{fig:IEC-A}b: arrows) left-shifted by $\sim28$~mT across the four samples. Meanwhile, the BL field segment shifts to the right, due to the ``reaction'' effective field on the BL from the ChL bilayer.

\paragraph{IEC-A: LTEM \& Sk Count}
To examine the effects of varying IEC-A on textural stability, we performed LTEM imaging along the $M(H)$ loop with decreasing field (\ref{fig:IEC-A}c). For $t_{\text{Pt-A}}=13$~Å (\ref{fig:IEC-A}c: right), skyrmions are formed at 10~mT, and elongate into stripes at ZF. Notably, this textural evolution shifts to ZF for $t_{\text{Pt-A}}=12$~Å (\ref{fig:IEC-A}c: centre) -- corresponding to optimal IEC-A -- where we demonstrate the stabilization of ZF skyrmions approximately 110~nm in size. Finally, for $t_{\text{Pt-A}}=11$~Å (\ref{fig:IEC-A}c: left), skyrmion formation shifts further to $-5$~mT. To distill essential microscopic information, we performed a manual skyrmion count for $5\times5$~$\mu$m$^{2}$ fields-of-view over a range of OP fields, and determined the skyrmion density evolution, $n_{\text{s}}(H)$, for four samples with varying $t_{\text{Pt-A}}$ (\ref{fig:IEC-A}d). In line with our previous work \citep{Chen.2022b}, for low IEC-A ($t_{\text{Pt-A}}=11$~Å: blue) $n_{\text{s}}(H)$ exhibits a dome shape that peaks near the saturation field, $H_{s}$. As $t_{\text{Pt-A}}$ is decreased, the $n_{\text{s}}(H)$ curve left-shifts by approximately $\sim23$~mT across the four samples, comparable to the left-shift of $M(H)$ in \ref{fig:IEC-A}b, while the peak value increases.

\paragraph{IEC-A: Discussion}
In \ref{fig:IEC-A}e, we plot the ZF skyrmion density ($n_{\text{s}}^{\text{Z}}$) thus obtained against $H_{\text{IEC-A}}$ (estimated from $t_{\text{Pt-A}}$ via \ref{fig:FuncLayers}h), which also yields a dome shape. Notably, our peak $n_{\text{s}}^{\text{Z}}$ of 7.1~$\mu$m$^{-2}$ is more than 3$\times$ that of a previous report using IEC \citep{Chen.2015} and also higher than that from exchange bias approaches \citep{Rana.2020}. We emphasize that while skyrmions form on both sides of the $t_{\text{Pt-A}}$ dome (\ref{fig:IEC-A}e), they are spatially isolated --- as required for racetrack applications --- only on the right side ($t_{\text{Pt-A}}\leq$12~Å), which therefore is the focus of our discussions hereafter. However, in this regime, $n_{\text{s}}^{\text{Z}}$ falls precipitously to near-zero ($7.1\rightarrow0.4$~$\mu$m$^{-2}$) within a very small range of $t_{\text{Pt-A}}$ ($12\rightarrow11$~Å) and $\mu_{0}H_{\text{IEC-A}}$ ($120\rightarrow150$~mT). As it stands, it is extremely challenging to reproduce and tune $n_{\text{s}}^{\text{Z}}$, which motivates the search for a secondary tuning mechanism.

\paragraph{IEC-B: MH Loop, LTEM, \& Sk Count}
To explore the role of tuning IEC-B, we varied $t_{\text{Pt-B}}$ from 11 to 6~Å, with fixed $t_{\text{Pt-A}}=12$~Å. The resulting $M(H)$ loops (\ref{fig:IEC-B}a,b) show the onset of ChL bilayer field evolution shifted rightwards for $t_{\text{Pt-B}}=11$ to 8~Å, and leftwards from 8 to 6~Å, crossing ZF. Notably, the total range of non-monotonic field-shift --- $\sim\pm$3~mT from ZF --- is an order of magnitude smaller than the monotonic shift obtained for varying IEC-A (\ref{fig:IEC-A}b). Intriguingly, despite the subtle nature of $M(H)$ shifts across $t_{\text{Pt-B}}$, LTEM reveals remarkable variation in imaged magnetic textures (\ref{fig:IEC-B}c). Notably, the ZF images (centre row) show the textural evolution transforming from sparse skyrmions to dense skyrmions, and then to stripes. The $n_{\text{s}}(H)$ curves (\ref{fig:IEC-B}d), similarly to $M(H)$, shift non-monotonically for $t_{\text{Pt-B}}=11$ to 6~Å, albeit with subtle differences in turning points ($n_{\text{s}}(H)$: 9~Å c.f. $M(H)$: 8~Å). Slight differences in $M(H)$ and $n_{\text{s}}(H)$ between nominally identical $t_{\text{Pt-B}}=$10~Å (\ref{fig:IEC-B}: red) and $t_{\text{Pt-A}}$=12~Å (\ref{fig:IEC-A}: green) reflects the extent of drift between sample depositions.

\begin{figure*}
	\centering \includegraphics[width=1\linewidth]{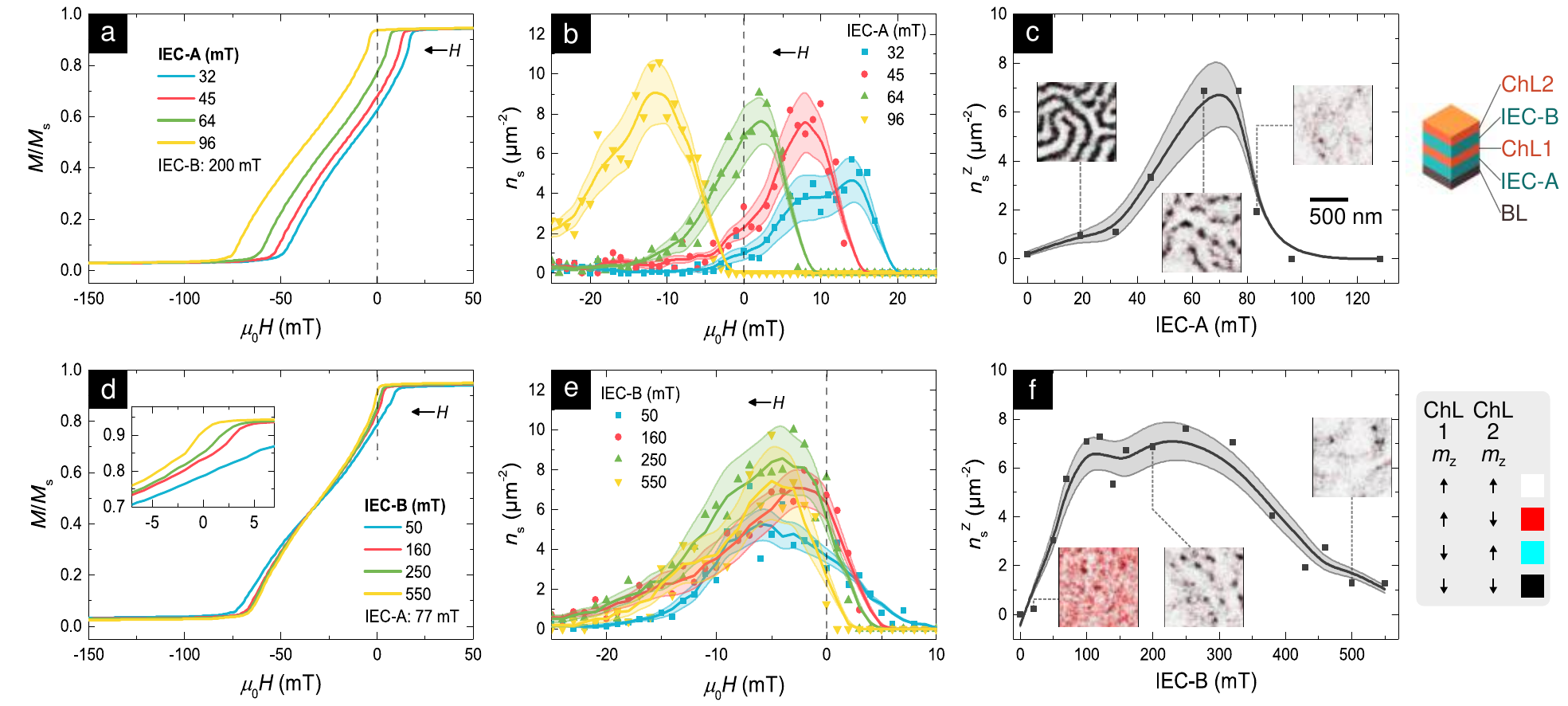} 
	\caption[Simulated Field Evolution of Textures]{\textbf{Micromagnetic Simulations of Field Evolution.}
	Simulated magnetization and textural field evolution trends for composite stack configurations (schematic on upper right) with \textbf{(a-c) }varying IEC-A ($\mu_{0}H_{\text{IEC-A}}$: $32\rightarrow96$~mT, $\mu_{0}H_{\text{IEC-B}}$ = 200~mT) and \textbf{(d-f) }varying IEC-B ($\mu_{0}H_{\text{IEC-A}}$ = 77~mT, $\mu_{0}H_{\text{IEC-B}}$: $50\rightarrow550$~mT). 
	(a, d) Simulated half $M(H)$ loops, focusing on the ChL bilayer evolution with decreasing $H$. Inset of (d) zooms-in to highlight the subtle changes with IEC-B.
	(b, e) Algorithmically computed field evolution of skyrmion density, $n_{\text{s}}(H)$. 
	(c, f) ZF skyrmion density, $n_{\text{s}}^{\text{Z}}$, as a function of $\mu_{0}H_{\text{IEC-A}}$ (c) and $\mu_{0}H_{\text{IEC-B}}$ (f) respectively. 
	Insets shows simulated ZF magnetization configurations generated by combining $m_{z}$ of ChL1 and ChL2 into a single image using different color channels. Color map (lower right) summarizes the representation methodology: ($\uparrow,\downarrow$ arrows depict the normalized OP magnetization, $m_{z}=\pm1$, for ChL1,2, while colored squares represent the assigned endpoints. Intermediate $m_{z}$ values are assigned interpolated colors. }
	\label{fig:Sims-Field}
\end{figure*}

\paragraph{IEC-B: Discussion}
Interestingly, $n_{\text{s}}^{\text{Z}}(H_{\text{IEC-B}})$ (\ref{fig:IEC-B}e) still follows a dome-shape (c.f. $n_{\text{s}}^{\text{Z}}(H_{\text{IEC-A}})$: \ref{fig:IEC-A}e), albeit with a much broader field extent due to the weaker dependence of $n_{\text{s}}(H)$ on IEC-B. This underscores the clear advantage of tuning $n_{\text{s}}^{\text{Z}}$ with IEC-B as compared with IEC-A: the drop in $n_{\text{s}}^{\text{Z}}$ from 7.4 to 0.4~$\mu$m$^{-2}$ as $t_{\text{Pt-B}}$ is reduced from 9 to 6~Å ($\mu_{0}H_{\text{IEC-B}}$: 240 to 430~mT) is 3$\times$ the $t_{\text{Pt}}$ range (6$\times$ in $\mu_{0}H_{\text{IEC}}$) obtained for varying IEC-A ($t_{\text{Pt-A}}$: 12 to 11~Å, $\mu_{0}H_{\text{IEC-A}}$: 120 to 150~mT). However, unlike the monotonic effect of IEC-A tuning, which stems from a shift in effective field, the physics underlying the non-monotonic IEC-B dependence of $M(H)$ and $n_{\text{s}}^{\text{Z}}$ on is unclear. In the following section, we use micromagnetic simulations to explore the possible microscopic mechanism at play.

\section{Micromagnetic Simulations\label{sec:Sims}}

\begin{figure}
	\centering \includegraphics[width=1\linewidth]{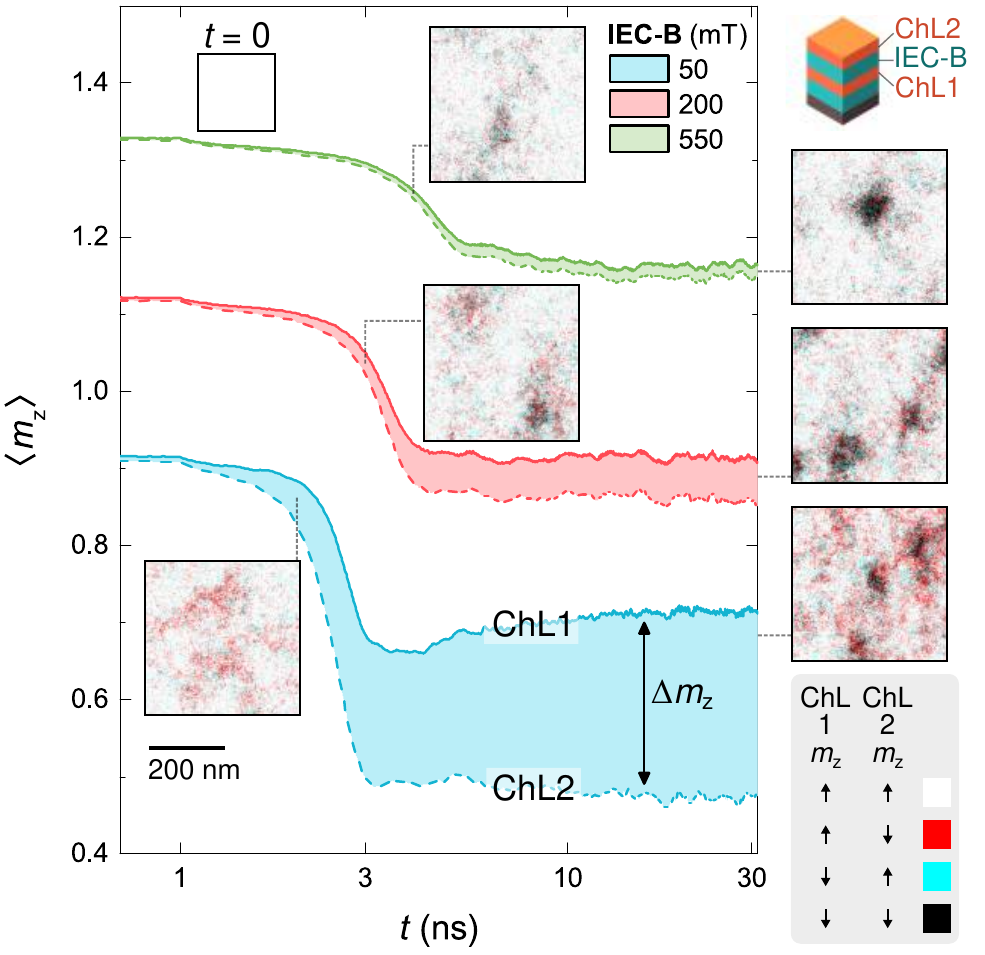}
	\caption[Transient Micromagnetic Simulations]{\textbf{Transient Micromagnetic Simulations.} 
	Time evolution of simulated magnetization for 3 stacks with varying $\mu_{0}H_{\text{IEC-B}}$ = 50 -- 550 ~mT (fixed $\mu_{0}H_{\text{IEC-A}}$ = 77~mT, $T=$600~K), with successive vertical offsets of 0.2 for clarity. The simulations are initialize with an uniform state with $m_{z}$=1. An external field (100~mT) is applied for the first 1~ns to maintain the uniform state, and then switched off. Upper solid (lower dashed) lines represent spatially-averaged OP magnetization , $\langle m_{z}\rangle$, of ChL1 (ChL2) respectively, and shaded region represents their difference, $\Delta m_{z}$. Inset images show magnetization configurations for each sample at two time slices --- domain nucleation (left) and end of simulation (right). The $m_{z}$ of ChL1,2 are combined using the methodology in \ref{fig:Sims-Field}, (colormap : lower right inset).}
	\label{fig:Sims-Transient}
\end{figure}

\paragraph{Simulations: Aim \& Validation}
Micromagnetic simulations can provide valuable spatial and temporal details that are not experimentally accessible \citep{Woo.2016,Buttner.2017,Legrand.2018,Woo.2018}. To validate our protocol, we first simulated hysteresis loops over 50 to $-150$~mT using previously established recipes \citep{Chen.2022} for stacks with varying IEC-A,B (details in Methods, SM §S4), and compared the $M(H)$ and $n_{\text{s}}(H)$ curves with experiments. With increasing IEC-A, the $M(H)$ curves rigidly left-shift by $\sim20$~mT across the four samples (\ref{fig:Sims-Field}a), in good agreement with experiments (\ref{fig:IEC-A}a). The skyrmion density, $n_{\text{s}}(H)$, (\ref{fig:Sims-Field}c) peaks $\sim10$~mT below $H_{s}$ and is likewise left-shifted, coupled with an increase in peak height. Consequently, the $n_{\text{s}}^{\text{Z}}$ peaks at $\mu_{0}H_{\text{IEC-A}}\sim60$~mT (\ref{fig:Sims-Field}e), in good agreement with experiments (\ref{fig:IEC-A}b,c). Here, we have combined the normalized OP magnetization, $m_{z}$ of ChL1 and ChL2 into a single image by using separate color channels to depict their relative magnitudes (\ref{fig:Sims-Field}e,f inset). This representation is especially useful for highlighting any decoupling between the two layers, which is relevant for subsequent discussions.

\paragraph{Simulations: IEC-B Effects}
In contrast to IEC-A, increasing IEC-B does not translate the $M(H)$ curves, but instead subtly compresses them in field (\ref{fig:Sims-Field}b: inset). This small but important trend is seen much more clearly in simulations compared to magnetometry experiments (\ref{fig:IEC-A}b). Expectedly, the same trend with IEC-B is also observed for the $n_{\text{s}}(H)$ curves (\ref{fig:Sims-Field}d). The subtle nature of this variation manifests as a broad peak in $n_{\text{s}}^{\text{Z}}$ with varying IEC-B (\ref{fig:Sims-Field}f), consistent with experiments (\ref{fig:IEC-B}e), proffering IEC-B as an ideal secondary knob for fine-tuning ZF skyrmion density.

\paragraph{Transient Sims Details}
Having experimentally anchored our simulations, we proceed to elucidate the role of IEC-B in skyrmion nucleation and stability. Starting from a uniformly magnetized state, we simulated the temporal evolution of ZF magnetization for three stacks with varying $\mu_{0}H_{\text{IEC-B}}$ (50--550~mT) and fixed $\mu_{0}H_{\text{IEC-A}}$ (77~mT) at 600~K. The elevated simulation temperature compensates for the short simulation time, and serves as an alternative to explicitly calculating energy barriers for skyrmion nucleation and annihilation \citep{Buttner.2018,CortesOrtuno.2017}. \ref{fig:Sims-Transient} shows the time evolution of $\langle m_{z}\rangle$, the spatially-averaged OP magnetization, for the three stacks. First, domain nucleation --- indicated by the decrease in $\langle m_{z}\rangle$ --- occurs more readily for lower IEC-B (\ref{fig:Sims-Transient}: blue curve). Simultaneously, the $\langle m_{z}\rangle$ of ChL1 and ChL2 begin to diverge, and $\Delta m_{z}$, the magnitude of divergence (\ref{fig:Sims-Transient}: shaded regions),varies inversely with IEC-B strength.

\paragraph{Transient Sims Insights}
Next, the simulated magnetization images (\ref{fig:Sims-Transient}: insets) show that for low IEC-B, the initial texture nucleation is limited to ChL2, while ChL1 is largely uniform (red textures in \ref{fig:Sims-Field}, bottom-left). Even at the end, the textures are still fairly decoupled across ChLs (red-black textures in \ref{fig:Sims-Field}, bottom-right; large $\Delta m_{z}$), which is undesirable for most applications \citep{Legrand.2018}. At the other extreme, a high IEC-B strongly couples the ChLs even in the nucleation phase, as reflected in the consistently small $\Delta m_{z}$ (\ref{fig:Sims-Transient}: green). This likely leads to a higher nucleation energy barrier, delayed nucleation, and reduced skyrmion densities. Crucially for the `Goldilocks' value of IEC-B ($\mu_{0}H_{\text{IEC-B}}=200$~mT), moderate decoupling is allowed between the ChLs, promoting the nucleation of considerable density of skyrmions, while ensuring their eventual coupling across both layers (black textures in \ref{fig:Sims-Field}, centre-right).

\begin{figure}
\centering \includegraphics[width=1\linewidth]{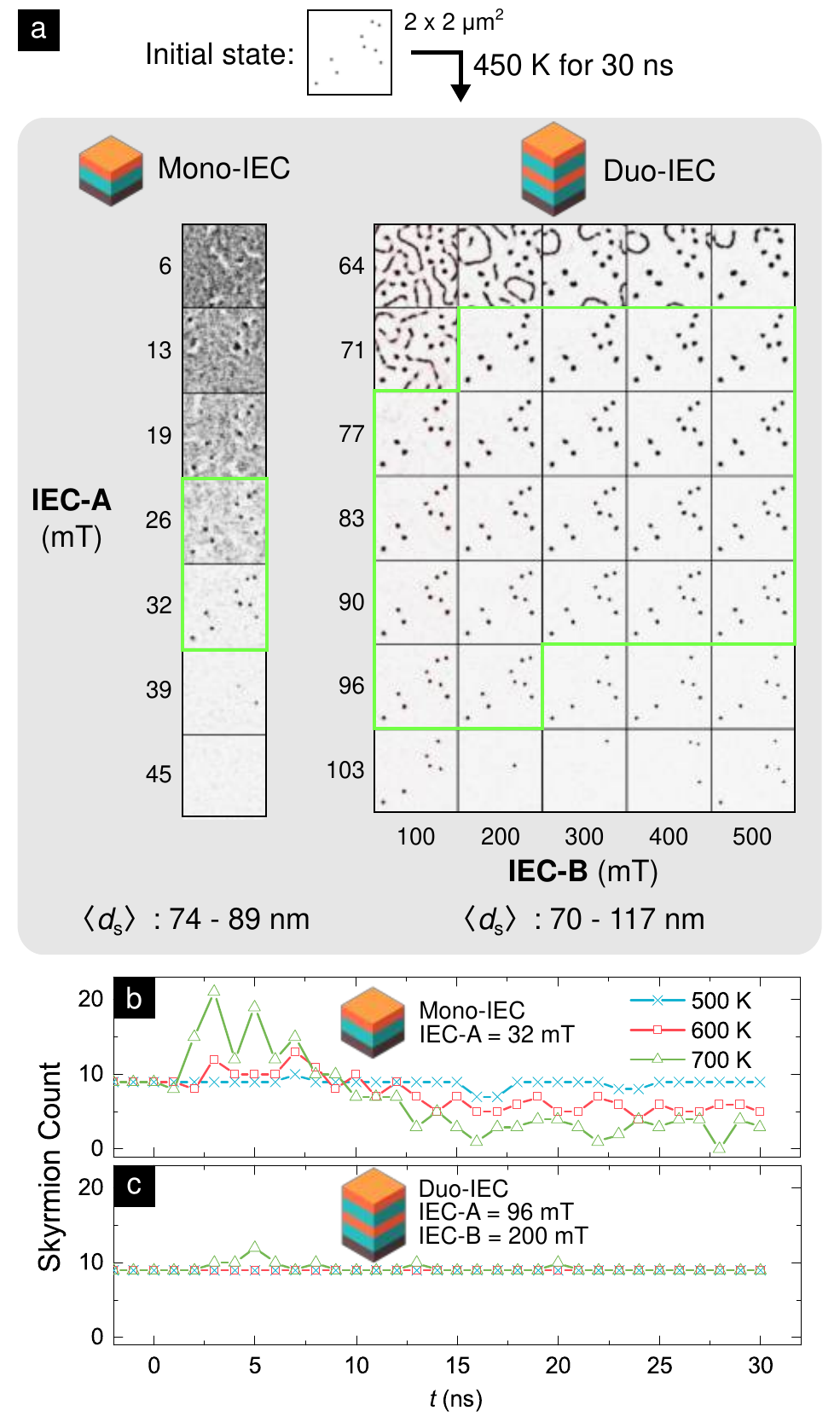}
\caption[Simulated Extent of Skyrmion Stability]{\textbf{Skyrmion Stability Simulations.}
\textbf{(a)} Simulated ZF magnetization for one (left) or two (right) ChLs coupled to a BL, for a range of IEC-A and IEC-B values (insets: stack structures). Top inset image shows the initialized state (nine skyrmions), using which each IEC set is simulated for 30~ns at 450~K. The final images combine the $m_{z}$ of ChL1 and ChL2 as before (method and colour map in \ref{fig:Sims-Field}). Green lines enclose the parameter space wherein skyrmions are stable against elongation and annihilation. The range of stable skyrmion diameters, $\langle d_{\text{S}}\rangle$ (averaged for each frame), is noted below. 
\textbf{(b,c)} Temporal evolution of the same initialized state (nine skyrmions), when subjected to varying high temperatures (500 -- 700~K) for single-IEC (b) and duo-IEC (c) stack structures. Simulations are initialized at 450~K, followed by elevated temperatures (500-700~K), while the skyrmion count is tracked over time.}
\label{fig:Sims-Stability}
\end{figure}

\paragraph{Textural Stability Sims}
In addition to its influence on skyrmion nucleation, we study the effect of IEC-B on textural stability, namely the ability to preserve existing magnetic textures without morphological alterations, e.g. skyrmion elongation, domain nucleation or annihilation. In \ref{fig:Sims-Stability}, we compare the final magnetization configurations for stacks with either one (mono-IEC) or two ChLs (duo-IEC) coupled to a BL with varying strengths. The simulations are initialized with nine isolated skyrmions (top inset), and run for 30~ns at 450~K (elevated to compensate for short simulation time). At low IECs (\ref{fig:Sims-Stability}a grid: top and left), the system is unstable to domain nucleation and skyrmion elongation, while, at high IECs (\ref{fig:Sims-Stability}a grid: bottom and right), some existing skyrmions are annihilated. Crucially, the parameter space of textural stability (\ref{fig:Sims-Stability}a grid: green borders) is substantially larger for duo-IEC than for mono-IEC. We further simulated the two stack structures at even higher temperatures (500 -- 700~K) (\ref{fig:Sims-Stability}b,c), and tracked the time evolution of skyrmion count as a proxy for textural stability. For the single-IEC stack (\ref{fig:Sims-Stability}b), the skyrmion count showed large fluctuations at temperatures above 500~K, and decayed rapidly with time. In contrast, for the duo-IEC stack (\ref{fig:Sims-Stability}c), the skyrmion counts remains very stable for all temperatures.

\paragraph{Sk Stability Sims Results}
These temporal simulations underscore the pivotal role of IEC-B in nucleating and stabilizing isolated ZF skyrmions. While the inclusion of IEC-B increases the minimum threshold for IEC-A (26 to 71~mT), the parameter space for skyrmion stability in two ChLs is considerably larger, with $\mu_{0}H_{\text{IEC-A}}$ ranging from 71 -- 96~mT (c.f. 26 -- 32~mT for one ChL). The range of suitable $\mu_{0}H_{\text{IEC-B}}$ is even greater (100 -- 500~mT). Moreover, the combination of two IECs supports a larger range of skyrmion sizes (\ref{fig:Sims-Stability}a: bottom), and greater tunability, than a single ChL. Notably, the duo-IEC system offers substantially larger textural stability against thermal fluctuations than the mono-IEC system (\ref{fig:Sims-Stability}b,c).

\section{Summary and Impact\label{sec:Conc}}

\paragraph{Results Summary}
In summary, we have demonstrated that the formation, density, and stability of ZF skyrmions in ultrathin chiral multilayers can be precisely tuned by exploiting a duet of IECs, each of which plays distinct physical roles. On one hand, the IEC coupling the ChL and BL introduces an effective field for the ChL, and thereby Zeeman-shifts the energetic stability of anti-aligned skyrmions, to enable their ZF stabilization \citep{Chen.2015,LoConte.2020,Rana.2020,Guang.2020}. On the other hand, the IEC coupling the two ChLs introduced in this work governs the dynamics of ZF skyrmion nucleation, likely by reshaping the nucleation energy barrier. Our experiments and simulations show that, when used in conjunction, IEC-A and B harmonize the tuning of ZF skyrmion density by modulating the energetics and kinetics of domain nucleation respectively. Crucially, a duo-IEC stack, hosting IECs with such complementary attributes, stabilizes ZF skyrmions over a much larger parameter space and temperature range as compared to mono-IEC stacks.

\paragraph{Impact \& Outlook}
The duo-IEC stack established through this work, shown to host isolated ZF skyrmions with tunable density and enhanced stability, is expected to find immediate use in device development efforts. First, its inherent compatibility with established stack requirements for electrical skyrmion generation and manipulation \citep{Litzius.2017,Jiang.2017,Finizio.2019,Woo.2018,Je.2021}, coupled with robustness to thickness variations, promises ease of device integration. Second, the demonstrated ability to precisely tune skyrmion density over a wide range via the thickness of IEC layers paves the way for application-specific stack designs to realize bespoke skyrmion configurations. Notably, the greatly enhanced ZF textural stability of the duo-IEC stack, with demonstrable robustness to thermal fluctuations, is highly relevant indevices such as skyrmionic artificial synapses \citep{Song.2020}, where skyrmions serve as the information carrier. Hence, our work opens up possibilities of using IEC engineering to host and manipulate skyrmions for next-generation computing applications.

\section{Methods}

\paragraph{Sample Fabrication and Characterization}
Multilayer films were deposited on pre-cleaned, thermally oxidized 200 mm Si wafers by ultrahigh vacuum magnetron sputtering using the Singulus Timaris$^{\text{TM}}$ system. For the CoB layer, a composite Co$_{80}$B$_{20}$ target was used. The Pt thickness, $t_{{\rm Pt}}$ is varied by wedged deposition of the relevant Pt layer across the wafer. Magnetometry results shown in \ref{fig:IEC-A} and \ref{fig:IEC-B} were obtained using a MicroMag Model 2900$^{\text{TM}}$ alternating gradient magnetometer in OP and IP orientation, while those in \ref{fig:FuncLayers} were obtained using a MagVision$^{\text{TM}}$ polar Kerr microscope.

\paragraph{Lorentz TEM Experiments}
For direct comparisons between magnetometry and microscopy, the films were simultaneously deposited on 20 nm-thick SiO$_{2}$ membrane grids from SPI Supplies. Lorentz transmission electron microscopy (LTEM) experiments were performed using an FEI Titan$^{\text{TM}}$ S/TEM operated in Fresnel mode at 300~kV. A dedicated Lorentz lens was used to focus the electron beam, with a defocus of -2.4~mm (see SM §S2). Meanwhile, the objective lens, located at the sample position, was switched off for field-free image acquisition, or energised to apply OP magnetic fields (-300~mT to +2~T) for \textit{in situ} studies of magnetic texture evolution. A finite tilt of 15° with respect to normal incidence was used to ensure that Néel textures exhibited sufficient domain contrast \citep{Chen.2022,Chen.2022b}. Custom-written Python scripts were used to perform background subtraction. Image analysis procedures are detailed in the SM (§S2).

\paragraph{Micromagnetic Simulations}
Micromagnetic simulations were performed using mumax$^{3}$ \citep{Vansteenkiste.2014}. The dimensions of the simulation cell were 4~nm $\times$ 4~nm $\times$ 3~nm, with the simulation space consisting of 512 $\times$ 512 $\times$ 4 cells. The effective medium approximation \citep{Woo.2016} was used to enable the modelling of IEC. The IEC strength, $J_{\text{IEC}}$, was obtained from $\mu_{0}H_{\text{IEC}}=J_{\text{IEC}}/(tM_{\text{s}})$, where $t$ is the thickness of the relevant FM layer. This was used to calculate the direct exchange scaling factor, $J_{\text{IEC}}t_{\text{cell}}/2A$, where $t_{\text{cell}}$ is the thickness of the simulation cell and $A$ is the direct exchange stiffness. Hysteresis loops were simulated using the protocol described in \citep{Vansteenkiste.2014}. The field range was limited to focus on the field evolution of the ChL bilayer for direct comparison with experimental results, while no domains were nucleated within the BLs.

\vspace{0.1ex}
\begin{center}
\rule[0.5ex]{0.5\columnwidth}{0.2pt}
\par\end{center}
\vspace{0.1ex}

\textsf{\textbf{\small{}Acknowledgments.}}{\small{}We acknowledge helpful experimental inputs from Felix Büttner, Simone Finizio, Kai Litzius, Hong Jing Chung, and Sze Ter Lim. This work was supported by the SpOT-LITE programme (Grant No. A18A6b0057), funded by Singapore's RIE2020 initiatives.}

\begin{center}
\rule[0.5ex]{0.4\columnwidth}{0.5pt}
\par\end{center}

\phantomsection
\def\bibsection{\section*{\refname}} 
\linespread{1.00}
\setlength{\parskip}{0.5ex}
\bibliographystyle{apsrev4-1}
\bibliography{Sk-IEC_Refs}

\begin{center}
\rule[0.5ex]{0.6\columnwidth}{0.5pt}
\par\end{center}

\end{document}